# Temperature reshapes epigenomic diversity in *Arabidopsis thaliana*: JSD and Methylator reveal RdDM–CMT2 plasticity

Eylül Gökce Harputluoglu [1,2,3], Ueli Grossniklaus [1,2,4], Deepak Kumar Tanwar [1,2,5,6]*

[1] Department of Plant and Microbiology, University of Zurich, Zollikerstrasse 107, 8008 Zurich, Switzerland.

[2] University Research Priority Program "Evolution in Action", University of Zurich, Zollikerstrasse 107, 8008 Zurich, Switzerland.

[3] Department of Informatics, University of Zurich, Binzmühlestrasse 14, CH-8050 Zürich, Switzerland.

[4] Zurich-Basel Plant Science Center, University of Zurich, ETH Zurich & University of Basel, Tannenstrasse 1, 8092 Zurich, Switzerland.

[5] Present Address: Swiss Institute of Bioinformatics, Winterthurerstrasse 190, 8057 Zurich, Switzerland.

[6] Present Address: Department of Quantitative Biomedicine, University of Zurich, Winterthurerstrasse 190, 8057 Zurich, Switzerland.

*Corresponding author: Deepak Kumar Tanwar, University of Zurich and Swiss Institute of Bioinformatics, Winterthurerstrasse 190, 8057 Zurich, Switzerland. E-mail: deepak.tanwar@uzh.ch

# Abstract

DNA methylation can be associated with phenotypic plasticity, yet how temperature shapes DNA methylation diversity in natural populations is unclear. Analyzing whole-genome bisulfite sequencing from 1075 *Arabidopsis thaliana* accessions grown at 10°C, 16°C, and 22°C, we quantified single-cytosine diversity using Jensen-Shannon Divergence (JSD). Diversity consistently peaked at intermediate methylation levels across the CpG, CHG, and CHH sequence contexts. Temperature modulated this diversity, primarily impacting intermediately methylated sites, with non-CG contexts (CHG and CHH) exhibiting increased diversity at warmer temperatures. Notably, at 22°C, CHH diversity patterns indicated altered balance between the RdDM and CMT2 pathways that regulate specific transposable element (TE) superfamilies. Furthermore, accessions from Southern Europe displayed higher non-CG diversity at 22°C compared to Northern European accessions. Our findings reveal that temperature influences the epigenomic diversity landscape, highlighting context-dependent plasticity, a dynamic interplay between silencing pathways, and potential geographic adaptation in response to environmental cues.

# Background

Epigenetic modifications, particularly DNA methylation, are fundamental regulators of genome function, influencing gene expression, transposon activity, and genome integrity [1]. Importantly, DNA methylation can stably change gene expression, affecting development and physiology. This layer of epigenetic gene regulation provides organisms, especially sessile plants where DNA methylation is often heritable, with a crucial mechanism for phenotypic plasticity, enabling them to respond and adapt to fluctuating environmental conditions. In plants like *Arabidopsis thaliana*, DNA methylation occurs in the CpG, CHG, and CHH sequence contexts [1]. While CpG methylation is maintained across cell divisions by a maintenance DNA methyltransferase similar to that in mammals [1], non-CG methylation (CHG and CHH) is regulated by distinct, dynamic pathways. CHH methylation is primarily established and maintained by the RNA-directed DNA methylation (RdDM) pathway involving the DRM2 methyltransferase [1], while both CHG and CHH methylation in heterochromatic regions, particularly transposable elements (TEs), are maintained by the CHROMOMETHYLASE 2 (CMT2) pathway, often linked to H3K9 dimethylation [5]. The interplay and environmental sensitivity of these non-CG pathways, especially RdDM and CMT2, are key to understanding the dynamic plant epigenome.

*A. thaliana*, with its extensive genetic diversity across natural populations and well-characterized genome, serves as an excellent model system for studying the interplay between environment, genetics, and epigenetics [2]. Large-scale projects have documented substantial natural variation in DNA methylation patterns among different *A. thaliana* accessions [2]. Spontaneous changes in methylation can also occur across generations, creating heritable epialleles that contribute to phenotypic variation independent of DNA sequence changes [3]. While the influence of environmental factors, such as temperature, on methylation levels is recognized [6], a quantitative understanding of how temperature shapes methylation diversity at single-cytosine resolution across broad natural populations is lacking. Investigating this diversity, rather than just average methylation levels, is critical for understanding the epigenetic heterogeneity within populations and its potential role in adaptive responses.

Quantifying epigenetic heterogeneity requires metrics sensitive to the distribution of methylation states across individuals. Jensen-Shannon Divergence (JSD), an information-theoretic measure, provides a robust, model-free approach to quantify the diversity or divergence of methylation patterns at specific genomic sites within a population [4]. JSD measures the increase in uncertainty when methylation states from multiple individuals are pooled, effectively capturing the variability beyond simple averages. This method has proven useful for identifying regions of conserved or diversified methylation in *A. thaliana* [4].

In this study, we utilize publicly available whole-genome bisulfite sequencing data from approximately 1,000 natural *A. thaliana* accessions grown under three controlled temperatures (10°C, 16°C, and 22°C) [2]. Applying JSD, we perform a population-scale analysis to investigate the impact of ambient temperature on single-cytosine methylation diversity. Our specific aims are to (i) characterize the genome-wide relationship between methylation level (MET) and methylation diversity (JSD) across the three temperatures; (ii) determine how temperature changes modulate JSD patterns within CpG, CHG, and CHH contexts, particularly in relation to genomic features like TEs; and (iii) explore how the activities of the CMT2 and RdDM pathways contribute to the observed temperature-responsive diversity landscape, particularly for non-CG methylation.

# Results and Discussion

## Conserved relationship between methylation-level and diversity is modulated by temperature

To understand how ambient temperature influences epigenetic variation across natural populations of *A. thaliana*, we first characterized the genome-wide relationship between methylation level (MET) and methylation Jensen-Shannon Divergence (JSD) using whole-genome bisulfite sequencing (WGBS) data from 1075 accessions grown at 10°C, 16°C, and 22°C. Consistent across all three temperatures and sequence contexts (CpG, CHG, CHH), we observed a distinct dome-shaped relationship where JSD, quantifying site-specific diversity, peaks at intermediate methylation levels (MET ≈ 0.2-0.8), shown in Figure 2a. This fundamental pattern indicates that the greatest heterogeneity among individuals occurs at sites exhibiting mosaic methylation, likely reflecting cellular or allelic variability where both methylated and unmethylated states coexist within the population. Sites with very low (LMC) or very high (HMC) methylation levels generally show low diversity, indicating conserved states across the population. The overall shape of this relationship remained remarkably stable across the 12°C temperature range, suggesting that the fundamental principles governing methylation variability are largely conserved.

While the overall relationship was conserved, temperature significantly modulated the magnitude of diversity, particularly at intermediately methylated sites. By calculating the per-site difference in diversity (ΔJSD) between warmer and cooler conditions, shown in Figure 2b, we found that temperature-driven changes predominantly occurred within the intermediate MET range (0.2-0.8), especially near the high-JSD ridge. Changes were minimal at highly methylated or unmethylated extremes. Importantly, the direction of change depended on the sequence context: non-CG contexts (CHG and CHH) consistently showed increased diversity (positive ΔJSD) at warmer temperatures (16°C vs 10°C, 22°C vs 10°C, and 22°C vs 16°C), indicating greater epigenetic heterogeneity under warmer conditions. This suggests that the pathways maintaining non-CG methylation, namely RdDM (primarily CHH) and CMT3/CMT2 (primarily CHG), may exhibit reduced fidelity or increased stochasticity at higher temperatures. Such an increase in maintenance error or dynamic activity could directly enhance cell-to-cell or individual-to-individual variation, potentially broadening the scope for phenotypic plasticity. In contrast, the CpG context, maintained by the highly stable MET1 pathway, showed a more mixed response with both increases and decreases in diversity within the intermediate methylation band, suggesting more complex, locus-specific modulation rather than a global trend.

## Geographic origin shapes Temperature-responsive epigenetic diversity

Recognizing that *A. thaliana* populations are adapted to diverse local climates, we investigated whether geographic origin influences temperature-responsive methylation diversity. Comparing accessions from Northern and Southern Europe grown at 22°C revealed differences, shown in Figure 2c. While the overall methylation-diversity landscape shape remained conserved, Southern European accessions displayed markedly higher non-CG (CHG and CHH) diversity compared to Northern accessions under this warm condition. This difference was again concentrated in the intermediate methylation range. Conversely, CpG diversity patterns were much more similar between the two populations. This finding suggests that epigenetic plasticity, particularly in non-CG contexts, may itself be an adaptive trait. Accessions native to warmer climates might have evolved methylation systems (RdDM and CMT2) that are inherently more dynamic or sensitive to thermal cues, potentially facilitating faster responses or maintaining a 'memory' of thermal stress. The relative stability of CpG diversity across geographic origins underscores the conserved, essential role of MET1-mediated maintenance for genome integrity, irrespective of local adaptation.

## Context-specific genomic localization of high diversity sites responds to temperature

We next examined the genomic distribution of the most diverse cytosines (Metastable Cytosines, MSC; MET 0.2-0.8, high JSD) across different features, shown in Figure 3a. The localization patterns starkly differed by context, reinforcing their distinct biological roles. CpG diversity was overwhelmingly concentrated in genic regions, primarily within introns (up to 71.5% at 22°C) and distal promoter regions (>1000 bp), together accounting for over 77% of CpG MSCs. This strongly links CpG variability to gene body methylation, where stochastic variation might influence transcriptional fine-tuning or splicing, providing regulatory flexibility without disrupting core gene function.

In contrast, non-CG diversity (CHG and CHH) was largely excluded from exons (≤1.1%) and enriched in heterochromatic and regulatory elements, consistent with roles in silencing and regulatory plasticity. CHG diversity was broadly distributed across introns, TEs, and promoter regions, showing relative stability across temperatures. The CHH context displayed the most dramatic temperature response. At 10°C and 16°C, CHH diversity was balanced across TEs (~29-31%), introns (~19-22%), and proximal promoter regions (~40% combined). However, at 22°C, a significant redistribution occurred: the proportion of CHH diversity within TEs increased notably (to 34.0%), while the contribution from proximal promoter regions (<1000 bp) decreased sharply (total from ~39% combined to ~29%), with introns remaining relatively stable. This suggests that warmer temperatures trigger a strategic re-allocation of CHH stochasticity. Increased diversity within TEs might reflect either reduced silencing fidelity or an active mobilization response prompting reinforcement via RdDM, potentially uncovering cryptic regulatory potential. Concurrently, the reduced diversity in proximal promoters could indicate a stabilization or canalization of methylation states at key regulatory elements to ensure a more robust, less variable transcriptional response to the warm environment. This temperature-induced remodeling highlights an adaptive interplay where CHH flexibility is concentrated in potentially adaptive regions (TEs) while being constrained at critical gene promoters.

Furthermore, the relationship between MET and JSD within these categories showed context-specific dynamics, Figure 3b. While low methylation (LMC) always displayed a strong positive correlation (higher MET means higher JSD), and high methylation (HMC) showed the expected negative correlation (higher MET means lower JSD, i.e., stability), the behavior differed at intermediate levels. Notably, for CHH MSC sites at 22°C, the correlation became significantly more positive. This paradoxical behavior suggests that under warm stress, even as methylation machinery (likely RdDM) attempts to increase methylation at these highly variable sites (increasing MET), the inherent instability or dynamic turnover prevents silencing, resulting in maximal population-level diversity (high JSD). This high-flux state might represent an adaptive strategy to maximize epigenetic variation in response to stress.

## Temperature modulates pathway activity and transposable element superfamily contributions, especially in CHH Context

Given the enrichment of non-CG diversity within TEs and its temperature sensitivity, we investigated the contributions of different TE superfamilies and the associated methylation pathways (CMT2 and RdDM), as shown in Figure 7.

Across all temperatures, CHG methylation was dominated by LTR/Gypsy elements, accounting for ~57-66% of methylated CHG sites. This aligns with the known preference of the CMT2 pathway, which robustly maintains CHG methylation in the bodies of these long heterochromatic TEs via a reinforcing loop with H3K9me2. The stability of this pattern across temperatures underscores the constitutive, temperature-resilient nature of CMT2-mediated silencing of core heterochromatin. Consistently, when TEs were classified by their primary regulatory pathway, Figure 7b, CMT2-targeted TEs accounted for the vast majority (~65-73%) of CHG methylation, peaking at 22°C, while RdDM-targeted TEs contributed less (~9-16%), decreasing at 22°C.

The CpG context also showed stability, with LTR/Gypsy being the major TE superfamily contributor (~32-40%), followed by RC/Helitron and DNA/MuDR elements. This reflects MET1's stable maintenance activity on heterochromatic repeats. A large fraction of CpG methylation occurred in TEs unassigned to either the CMT2 or RdDM pathways (~56-63%), highlighting MET1's broad role beyond the specific targets of non-CG pathways.

The CHH context revealed the most significant temperature-dependent plasticity, Figure 7a. At 10°C and 16°C, RC/Helitron TEs were the primary source of CHH methylation (~36-39%). Accordingly, RdDM-targeted TEs dominated CHH methylation (~52%) under these cooler conditions, Figure 7b, consistent with RdDM's role in active de novo silencing. However, at 22°C, a striking shift occurred: the contribution from RC/Helitron decreased (to 21.6%), while the contribution from the DNA superfamily surged dramatically (to 27.4%), becoming the largest single superfamily fraction. Concurrently, the overall contribution of RdDM-targeted TEs decreased (to 38.3%), while that of CMT2-targeted TEs increased substantially, becoming the largest fraction (39.4%). This inverse dynamic strongly suggests a temperature-induced shift in the balance between RdDM and CMT2 activities in the CHH context. Elevated temperatures, potentially acting as a stressor that reactivates certain TEs, appear to attenuate RdDM activity or alter its targets, leading to a compensatory increase in the contribution of the CMT2 pathway (which also mediates some CHH methylation in deep heterochromatin). This reveals a plastic interplay where silencing pathways dynamically adjust their roles to maintain genome integrity under varying environmental conditions.

## Transposable element body and boundary dynamics reveal pathway-specific temperature responses

Analyzing the spatial distribution of methylation (MET) and diversity (JSD) across TE bodies and their flanking regions provided further insights into pathway-specific temperature responses, as shown in Figure 4 and Figure 5.

In the CpG context, CMT2-targeted TEs (especially LTR/Gypsy) showed constitutively high methylation within the TE body across all temperatures, consistent with robust MET1 activity. JSD profiles for these elements were also stable, often exhibiting distinct peaks or "shoulders" just outside the TE boundaries. These boundary JSD peaks likely mark the interface where silencing machinery actively polices TE borders to prevent heterochromatin spreading, representing consistent hotspots of epigenetic variability. RdDM-targeted TEs consistently showed low CpG methylation and JSD.

Non-CG contexts revealed clear temperature sensitivity. In the CHG context, CMT2-targeted LTR/Gypsy elements maintained high methylation within the TE body across temperatures, with average JSD peaking at 22°C. In contrast, RdDM-targeted TEs showed high CHG methylation and JSD at 10°C and 16°C, but both signals were substantially reduced at 22°C, reinforcing the heat sensitivity of RdDM.

In the CHH context, CMT2-targeted TEs (especially DNA and DNA/MuDR superfamilies) exhibited low methylation and JSD at 10°C and 16°C but showed a sharp, substantial increase in both MET and JSD across the TE body and flanks specifically at 22°C. Conversely, RdDM-targeted TEs (including RC/Helitron and LTR/Gypsy) displayed high methylation and sharp JSD peaks at their boundaries (a signature of RdDM border reinforcement) at 10°C and 16°C, but both signals strongly diminished at 22°C. Heatmaps of individual metastable TEs confirmed these opposing trends, showing distinct clusters dominated by specific superfamilies exhibiting either a gain (CMT2-targeted DNA/MuDR in CHH) or loss (RdDM-targeted RC/Helitron, LTR/Copia in non-CG) of methylation/JSD at 22°C, Figure 5.

This inverse dynamic between RdDM and CMT2 targets in the CHH context at 22°C can be a key finding. It strongly supports the model of a plastic, adaptive epigenetic defense system. When the primary de novo silencing pathway (RdDM) appears compromised or less effective under warmer conditions, the CMT2 pathway seems to be engaged more strongly, potentially compensating by reinforcing silencing on a different set of TE targets (primarily DNA/MuDR). This highlights a sophisticated interplay and potential crosstalk between the major non-CG methylation pathways, enabling the plant to maintain overall genome silencing across a range of temperatures by dynamically reconfiguring pathway activities. The attenuation of RdDM activity under heat stress aligns with previous studies showing environmental stress can compromise RdDM-mediated silencing. The compensatory increase in CMT2-associated CHH methylation reveals a previously underappreciated layer of epigenetic responsiveness to environmental cues.

# Methods

## Data source and plant growth conditions

Whole-genome bisulfite sequencing (WGBS) data for this study were sourced from the public dataset deposited in the NCBI Gene Expression Omnibus (GEO) under accession number GSE80744 [10]. This dataset, part of the *Arabidopsis thaliana* 1001 Epigenomes Project, includes methylomes for 1,107 natural accessions originally curated by the 1001 Genomes Project [9, 10].

As described in the original experiment [10], the plant growth protocol was standardized for all accessions. Seeds were stratified at 4°C for 3 days to ensure uniform germination. Following stratification, plants were grown for 4 weeks in a controlled environment under a 16h light / 8h dark photoperiod. The sole experimental variable was ambient temperature; cohorts of each accession were grown in parallel at constant temperatures of 10°C, 16°C, or 22°C. For all samples, rosette leaves were harvested for genomic DNA extraction. The original sequencing was performed on Illumina HiSeq 2500 and Illumina HiSeq 4000 platforms [10].

## DNA methylation data analysis workflow

Raw sequencing data (FASTQ) were downloaded from the Sequence Read Archive (SRA) for all samples associated with GSE80744. We processed this data using the Methylator framework [13], which provides a standardized pipeline for quality control, alignment, and methylation calling.

From the 1,107 available accessions, 1,075 accessions were selected for the final analysis based on post-processing quality metrics. The analysis workflow involved several key stages. Raw sequencing reads were first assessed for quality using FastQC. Reads then underwent preprocessing with TrimGalore [84] to filter low-quality (Phred score < 30) bases, trim adapters, and remove undetected bases (N). Subsequently, Clumpify was used to remove duplicate reads and merge data from different sequencing runs. Processed reads were aligned to the *A. thaliana*

TAIR10 reference genome using Bismark [85] implementing the ‘Dirty-Harry’ method [13]; this protocol performs a stringent end-to-end alignment followed by a local alignment for any remaining unmapped reads. Finally, methylation calls were extracted from both alignment steps and merged to create a comprehensive set of methylation calls for analysis.

## Quantification of methylome diversity and divergence

To quantify epigenetic diversity at single-cytosine resolution across populations, we adopted the information-theoretic approach, Jensen-Shannon Divergence (JSD), as previously described [1]. JSD is a model-free, non-parametric metric that quantifies the dissimilarity between a set of probability distributions.

In the context of methylome data, each cytosine site is associated with a set of $j$ probability distributions $(Pj)$, where each distribution represents the methylation state (i.e., the counts of methylated vs. unmethylated reads) in a single sample $j$. JSD measures the increase in uncertainty (or, equivalently, the loss of information) that occurs when these distinct sample distributions are pooled into an average "mixture" distribution $\langle P \rangle$.

JSD is formally defined as the difference between the Shannon entropy H of the mixture distribution and the weighted average entropy $\langle H \rangle$ of the individual distributions:

$$JSD(P) = H\ (j \sum \pi jPj\ ) - j \sum \pi jH(Pj) = H(\langle P \rangle) - \langle H \rangle$$

Where $\pi j$ is the weight of each sample j. The Shannon entropy H for any discrete distribution Pj with states k (in this case, k=1 for methylated and k=2 for unmethylated) is defined as:

$$H(Pj) = -\ k \sum Pjk \log 2\ Pjk$$

We computed JSD using the "plug-in" estimator, which replaces the true probabilities Pjk with the observed frequencies from the read counts [1]. For a given cytosine site i, let nijk be the read count for methylation state k in sample j. The weight for each sample $\pi$^ij is its sequencing coverage $nij=\sum knijk$ relative to the total coverage $ni=\sum jnij$ at that specific site [1]:

$$\pi \hat{}ij=ninij$$

The resulting JSD value is measured in bits, with a theoretical range of 0 to 1 for a binary system. A JSD of 0 indicates perfect conservation, whereas a JSD of 1 indicates maximum divergence.

### Weighted methylation level calculation

As a complementary measure to JSD, we calculated the weighted average methylation level (MET) for each cytosine site i. MET represents the plug-in estimate of the methylation bias across the entire population and is calculated as the total number of methylated reads (k=1) across all samples (j) divided by the total coverage at that site:

μ^i=ni∑jnij1

Together, the JSD and MET values for each cytosine define a "phase plane" (Figure 1c) that characterizes its methylation state and diversity across the population.

## Population stratification and statistical analysis

The final dataset, comprising 1075 accessions, each with data from three temperature conditions, was stratified into distinct groups to perform comparative JSD and MET analyses. Calculations were performed independently for each of the three cytosine sequence contexts (CpG, CHG, and CHH). The primary analytical groupings included:

- Temperature (10°C, 16°C, or 22°C)
- Country of Origin
- Geographical Area
- Accession
- Interaction Terms (e.g., Geographical Area + Temperature)

All diversity calculations were performed using our R Bioconductor package, shannonR [11]. All downstream statistical analyses and data visualizations were conducted in R.

# Conclusion

In this study, we employed Jensen-Shannon Divergence (JSD) to conduct a large-scale, population-level analysis of single-cytosine methylation diversity in *Arabidopsis thaliana* under varying ambient temperatures. Our findings demonstrate that while the fundamental relationship between methylation level and diversity is conserved, temperature is associated with differences in diversity, primarily at intermediately methylated sites, under our 4-week rosette leaf sampling regime. We observed a pronounced, context-dependent effect, with non-CG methylation (CHG and CHH) exhibiting increased diversity under warmer conditions (22°C). This temperature-responsive plasticity was strongly associated with transposable elements (TEs).

Importantly, our analysis revealed evidence for a dynamic interplay between the RdDM and CMT2 pathways in shaping the CHH methylation landscape at 22°C. The observed decrease in methylation and diversity at canonical RdDM target TEs, coupled with a concomitant increase at CMT2 target TEs (particularly DNA transposons), suggests a compensatory mechanism where the epigenetic silencing machinery adapts to thermal stress by rebalancing pathway contributions. This highlights a previously underappreciated plasticity within the plant's genome defense systems.

Furthermore, the observation that accessions from warmer native climates (Southern Europe) display inherently higher non-CpG diversity at 22°C suggests that epigenetic responsiveness itself may be subject to local adaptation. The stable conservation of CpG methylation patterns, both across temperatures and geographic origins, underscores its fundamental role in maintaining genome integrity.

Taken together, our results establish that ambient temperature significantly sculpts the landscape of epigenetic diversity in natural plant populations. This environmentally modulated heterogeneity, particularly within the dynamic non-CG contexts and concentrated at TEs, represents a potentially crucial layer of variation. It may facilitate phenotypic plasticity and provide novel substrates for adaptation in response to changing environmental conditions. Understanding the mechanisms and consequences of this epigenetic diversity is essential for a complete picture of how plants adapt and evolve.

# Authors' contributions

DKT led the study's conceptualization. UG and DKT secured funding. EGH and DKT developed the methodology and conducted the formal data analysis. EGH was responsible for software development. UG provided resources and oversaw project administration. UG and DKT supervised the research. EGH, UG and DKT interpreted the data. EGH prepared the original manuscript draft, and UG and DKT contributed to the review and editing of the manuscript.

# Ethics declarations

## Ethics approval and consent to participate

Not applicable.

## Consent for publication

All authors have consented to the publication of this work in Genome Biology.

## Competing interests

None declared.

# Funding

This project was supported by the University Research Priority Program “Evolution in Action” and core funding from the University of Zurich to UG.

# Acknowledgments

Thank you to all UG lab members for insightful discussions. We also gratefully acknowledge Prof. Jürgen Bernard for officially supervising EGH during her master thesis. We also thank Prof. Mark Robinson for valuable feedback and helpful discussions on R package.

# Availability of data and materials

## Data availability

The genomic and methylome data for the *Arabidopsis thaliana* samples were sourced from the 1001 Genomes Project resource and its related public repositories.

## Code availability

The custom R package used for the analysis, shannonR, is publicly available on GitHub at https://github.com/urppeia/shannonR. Data analysis code is available on GitHub at https://github.com/urppeia/meth1001_athaliana

# Supplementary Information

Supplementary figures and tables are provided with the manuscript.

# Figure captions

## Figure 1

**Graphical overview of the study design and analytical framework. (a)** Global distribution of the 1075 *Arabidopsis thaliana* accessions used in the study, with collection sites across Europe, North America, Asia, and other regions. **(b)** The Shannon entropy formula underpinning Jensen-Shannon Divergence (JSD), the metric used to quantify single-cytosine methylation diversity across individuals. **(c)** Conceptual relationship between average methylation level (MET) and JSD, showing a characteristic dome-shaped pattern with peak diversity at intermediate MET values (≈0.2–0.8). Metastable cytosines (MSCs) are defined within this high-diversity region (red box). Inset illustrates methylation state distributions across individuals for low (LMC), medium (MMC), and high (HMC) methylation categories, highlighting the source of JSD variation.

## Figure 2: Methylation-Diversity relationship across Temperatures

### 2a.Methylation-Diversity relationship across temperature.

Each panel shows per-site methylation level (x-axis, 0–1) versus epigenomic diversity quantified by Jensen Shannon divergence (JSD, y-axis, bits, 0–1). Columns correspond to temperatures (10 °C, 16 °C, 22 °C); rows to sequence contexts (CHG, CHH, CpG). Points are hex-binned; with the fill indicating log10 density (purple to orange). Blue dashed vertical lines at 0.2 and 0.8 delineate low (LMC) and high (HMC) methylation categories, respectively. The intermediate band (0.2–0.8) is further partitioned into intermediate methylation (MMC, low diversity) and moderate methylation (MSC, high diversity). Panel headers report Spearman's ρ between methylation and JSD, associated p-value, and the number of sites (n).

### 2b. Difference (ΔJSD) plot.

For each genomic site, we computed the per-site difference in diversity between temperatures, ΔJSD = JSD (warmer) − JSD (cooler). Panels are arranged by contrast (columns: 16 °C–10 °C, 22 °C–10 °C, 22 °C–16 °C) and context (rows: CHG, CHH, CpG). Points are hex-binned in the methylation–JSD plane (x-axis: methylation level; y-axis: JSD at the warmer condition in the contrast), and fill encodes ΔJSD with a diverging scale (blue < 0, white ≈ 0, red > 0). The blue dashed box reproduces the medium-methylation (x = 0.2–0.8) and diversity range to orient comparisons; it does not indicate a ΔJSD threshold.

### 2c. Difference plot at 22°C between Northern Europe and Southern Europe.

**Left**: Per-site diversity difference ΔJSD between Northern and Southern Europe at 22°C. Points are plotted by methylation level (x, 0–1); color encodes ΔJSD on a diverging scale (blue < 0, white ≈ 0, red > 0). The blue dashed box marks the mid-methylation interval (0.2–0.8) and a reference diversity band for orientation. **Right**: JSD (bits) versus methylation at 22°C for Northern and Southern Europe (two columns) across CHG/CHH/CpG (rows). Hex-binned density; dashed verticals at 0.2 and 0.8 and MMC/MSC guides are shown. Panel headers report Spearman's ρ, p-value, and sample count (n).

## Figure 3: Genomic localization and chromosome-level correlation of high-diversity cytosines (MSCs) across contexts and temperatures

### 3a. Distribution of High-Diversity Cytosines (MSC) Across Genomic Features by Context and Temperature

This bar plot illustrates the percentage distribution of cytosines classified as Medium Methylation/High Diversity (MSC) (Methylation 0.2< x <0.8, High JSD) across various genomic features. Data are segregated by sequence context (CHG, CHH, and CpG) and presented for each temperature condition (10C, 16 C, and 22 C). The genomic features analyzed are Exonic, Intronic, Intergenic, two Promoter regions (<1000 bp and >1000 bp), Transposable Elements (TE), and the Transcription Start Site (TSS). Bar height represents the proportion of total MSC sites found within that specific feature for the given context and temperature.

### 3b. Correlation between MET and JSD

The plot shows the Spearman correlation coefficient (ρ) of JSD and the mean methylation level (MET) across the five *Arabidopsis thaliana* chromosomes (Chr1–Chr5). The analysis is performed across three segregation axes: sequence context (CpG, CHG, CHH), temperature condition (10°C, 16°C, 22°C), and a stratification by methylation-JSD categories (Low, Medium, and High MET and metastable cytosines, MSCs). The coefficient ρ defines the strength and direction of the monotonic relationship between MET and JSD, revealing how this fundamental stability-diversity trade-off is modulated by genomic location and environmental conditions.

## Figure 4: TE boundaries exhibit class-specific and temperature-dependent methylation and divergence dynamics

Enriched heatmaps and average profiles of methylation and JSD across TE regions and 2 kb flanking sequences. The data are partitioned by cytosine context types and further stratified by the TE's target silencing pathway (CMT2 or RdDM) and its superfamily. The line plots above each heatmap show the mean MET and JSD values for all TEs in that category. The heatmaps display the values for individual TEs (rows), where red indicates high values (approaching 1) and blue indicates low values (approaching 0). Profiles are shown for each of the three ambient growth temperatures (10°C, 16°C, and 22°C), revealing distinct patterns within TE bodies and at their boundaries.

## Figure 5: Specific heterochromatic TE superfamilies show distinct temperature-responsive methylation and divergence profiles

Heatmaps displaying normalized methylation levels (Meth) and JSD for individual metastable transposable elements (TEs). Rows represent individual TEs and are ordered by their first principal component score based on their methylation and JSD values across temperatures, highlighting the dominant axes of variation. The TEs are partitioned by cytosine context (CpG, CHG, CHH) and their association with the CMT2 or RdDM silencing pathways, or are unassigned. Major clusters of TEs are annotated by their superfamily. Columns represent the three ambient growth temperatures (10°C, 16°C, and 22°C). Color intensity corresponds to the value of MET or JSD, with dark red indicating high values (1) and light yellow indicating low values (0). Grey indicates missing data for a specific TE at a given temperature.

## Figure 6: Transposable element superfamily composition and pathway assignment of methylated cytosines across contexts and temperatures

### 6a. Contribution of transposable element superfamilies across contexts and temperatures

The bar charts display the percentage of methylated cytosines that fall within annotated TE superfamilies. Data are categorized by the three major cytosine sequence contexts (CHG, CHH, and CpG; columns) and by the ambient growth temperature of the *Arabidopsis thaliana* plants (10°C, 16°C, and 22°C; rows). Each bar represents the proportion of total methylation for a given condition that is associated with a specific TE superfamily.

### 6b. Contributions of CMT2 and RdDM pathways to the transposable elements

The bar charts illustrate the percentage of methylated cytosines that are located within transposable elements (TEs) targeted by distinct methylation pathways. The TEs are categorized as being predominantly targeted by the CMT2 pathway, the RdDM pathway, or are unassigned. Data are stratified by cytosine sequence context (CHG, CHH, and CpG; columns) and by the ambient growth temperature of *A. thaliana* plants (10°C, 16°C, and 22°C; rows). Each bar represents the percentage of total methylation for a given condition that is attributed to one of these three TE classes.

## Figure 7: MethDivergence pipeline for genome-wide quantification of epigenetic divergence

Overview of the computational workflow for calculating Jensen-Shannon Divergence (JSD) from methylation data. The pipeline is composed of several key modules. (Top Left) Input Data: The pipeline requires two inputs: Tabix-indexed BED files containing per-cytosine methylation counts and a metadata file linking sample IDs to their respective file paths. (Top Center) Processing Pipeline: The core workflow retrieves genomic regions, extracts methylation data, and applies quality control filters, including a minimum sample size and read count per site. Processing is performed in parallelized chunks for computational efficiency. (Top Right) Mathematical Framework: The pipeline's core calculation is the Jensen-Shannon Divergence (JSD), which is derived from Shannon Entropy. Probabilities are calculated from the methylated and unmethylated read counts at each cytosine position. (Center) Output Results: The primary output is a tab-separated text file providing per-position metrics, including genomic coordinates, JSD, sample size, read counts, and the weighted methylation level (MET). (Bottom) Structure, Specifications, and Implementation: The pipeline is implemented as an R package with a modular structure. An example code snippet demonstrates its straightforward implementation for analyzing a set of samples.

# Figure 1

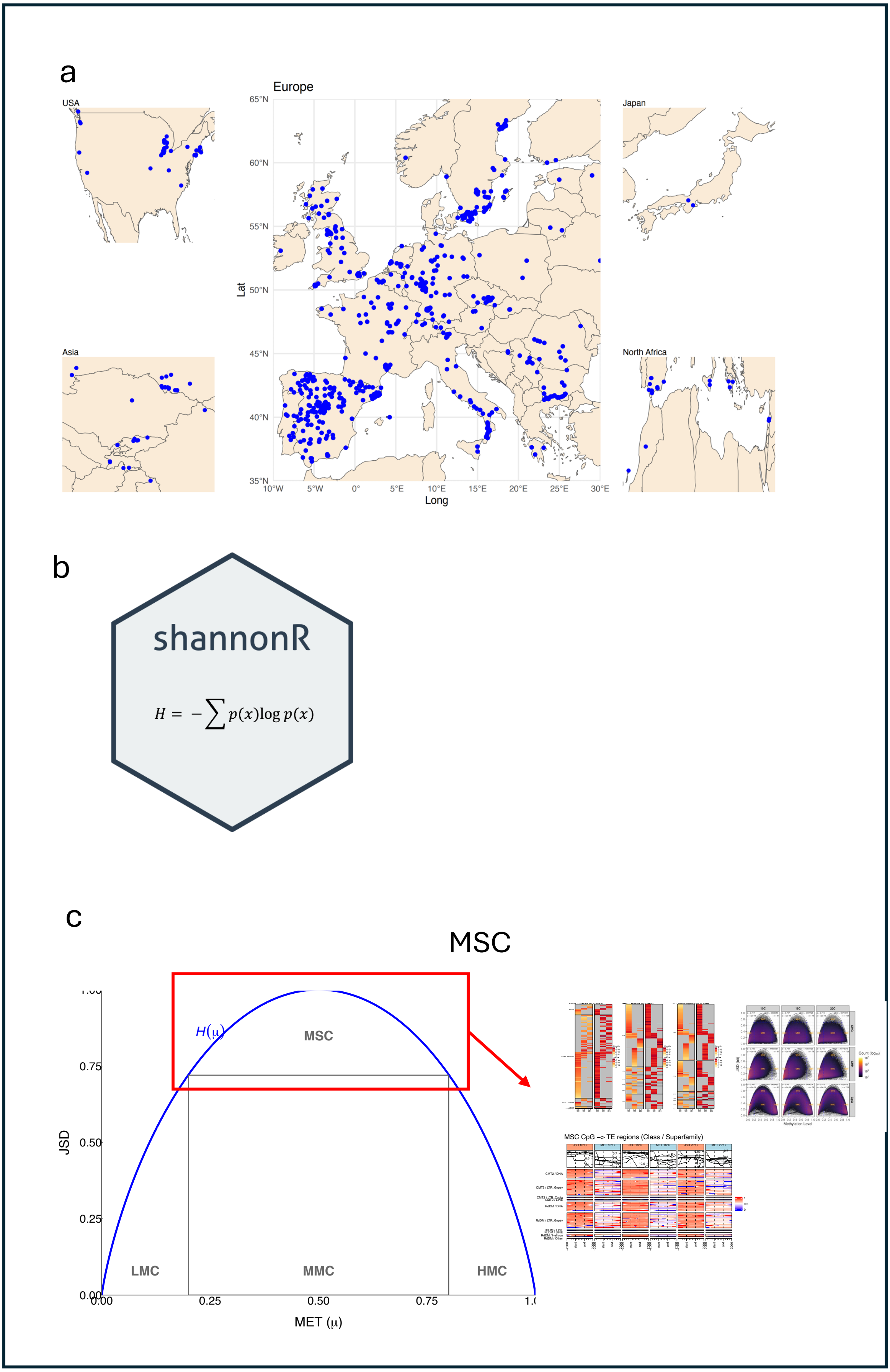
a
USA
Europe
Japan
Asia
North Africa
65°N
60°N
55°N
50°N
45°N
40°N
35°N
Lat
10°W
5°W
0°
5°E
10°E
15°E
20°E
25°E
30°E
Long
b
shannonR
$H = -\sum p(x) \log p(x)$
c
MSC
H(μ)
MSC
LMC
MMC
HMC
0.75
0.50
0.25
0.00
JSD
0.00
0.25
0.50
0.75
MET (μ)
MSC CpG --> TE regions (Class / Superfamily)

Figure 2

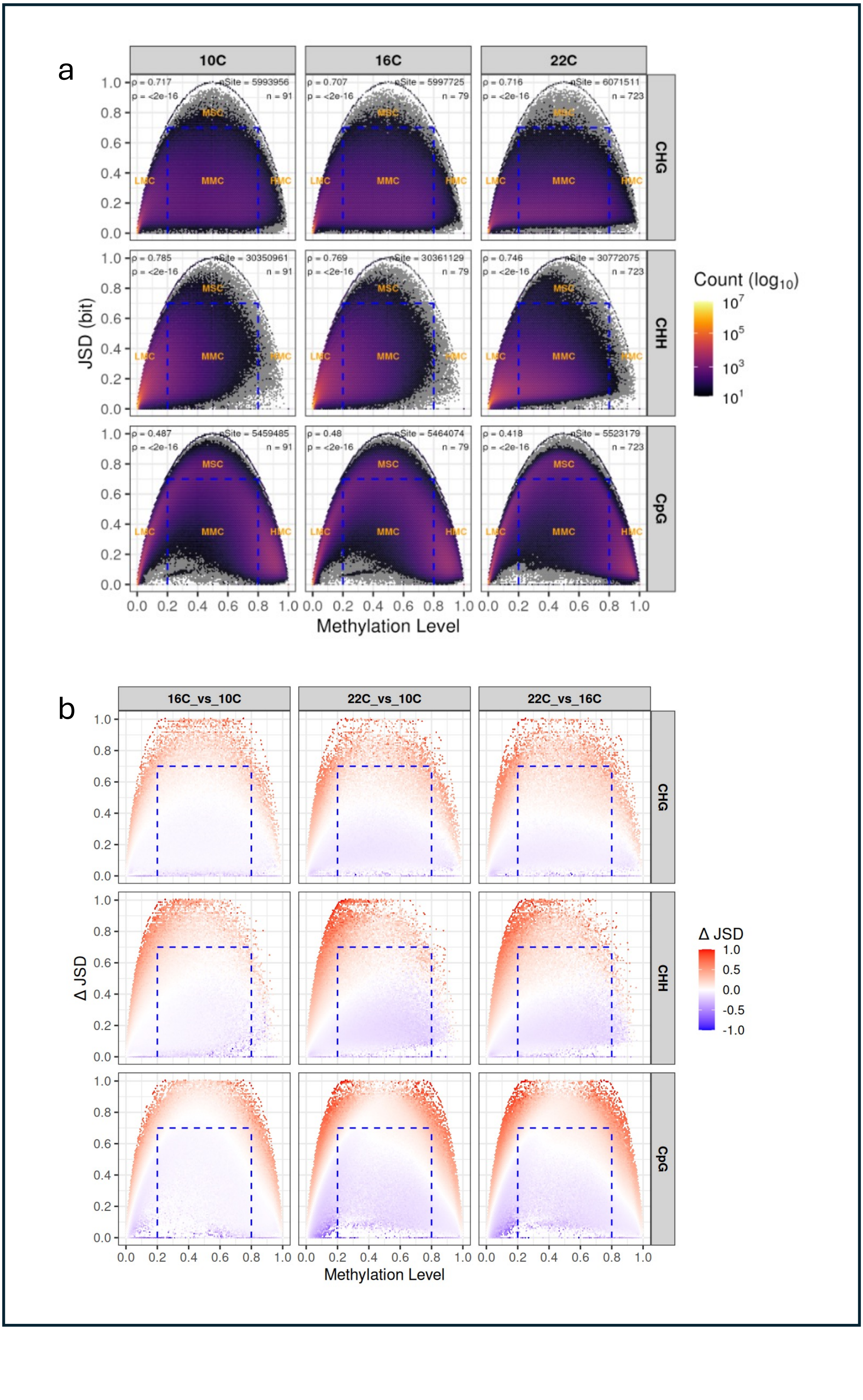
a
10C
16C
22C
CHG
CHH
CpG
ρ = 0.717 p = <2e-16 nSite = 5993956 n = 91
ρ = 0.707 p = <2e-16 nSite = 5997725 n = 79
ρ = 0.716 p = <2e-16 nSite = 6071511 n = 723
ρ = 0.785 p = <2e-16 nSite = 30350961 n = 91
ρ = 0.769 p = <2e-16 nSite = 30361129 n = 79
ρ = 0.746 p = <2e-16 nSite = 30772075 n = 723
ρ = 0.487 p = <2e-16 nSite = 5459485 n = 91
ρ = 0.48 p = <2e-16 nSite = 5464074 n = 79
ρ = 0.418 p = <2e-16 nSite = 5523179 n = 723
MSC
LMC
MMC
HMC
JSD (bit)
Methylation Level
Count (log10)
10^7
10^5
10^3
10^1
b
16C_vs_10C
22C_vs_10C
22C_vs_16C
Δ JSD
1.0
0.5
0.0
-0.5
-1.0
Methylation Level

C

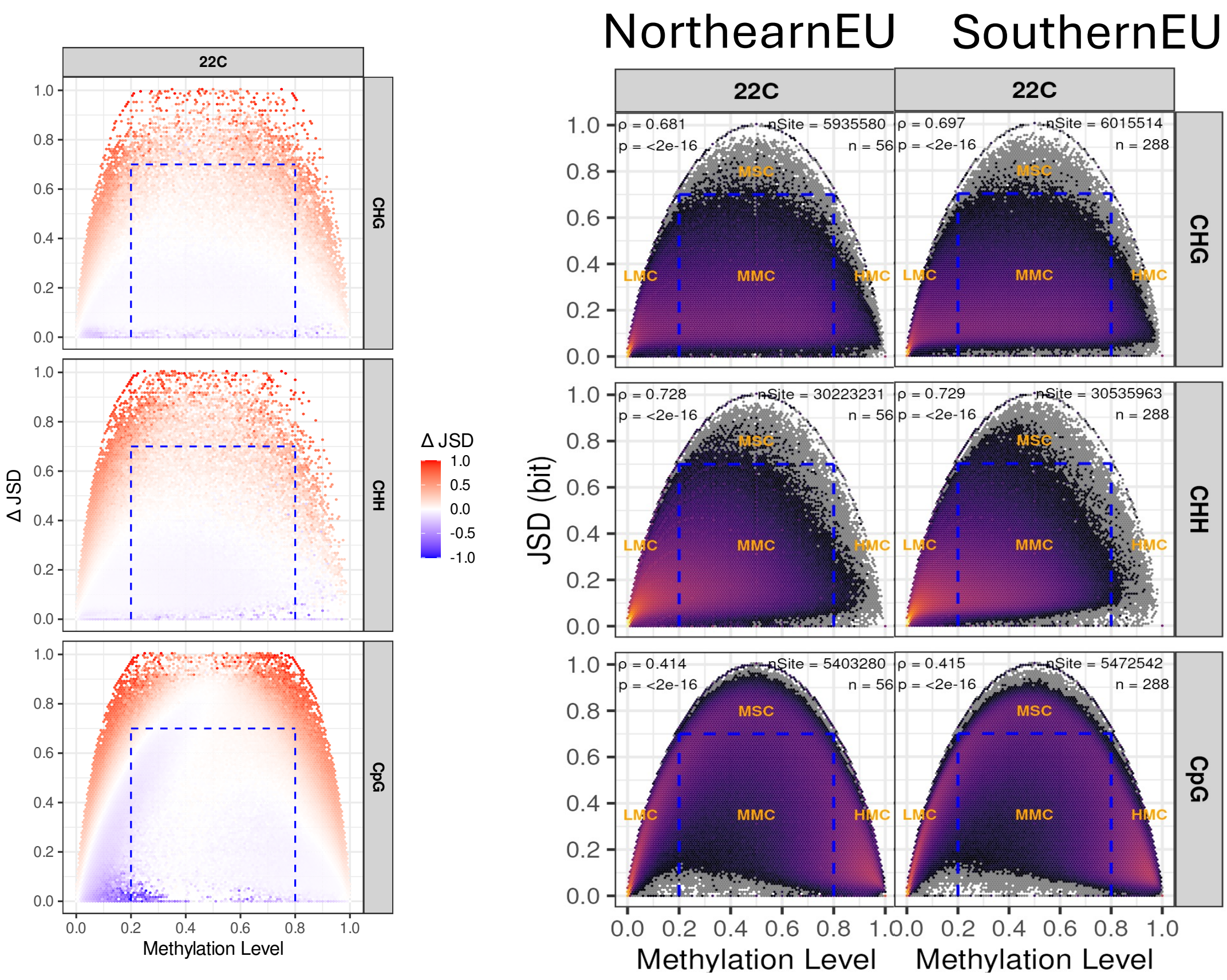

Figure 3

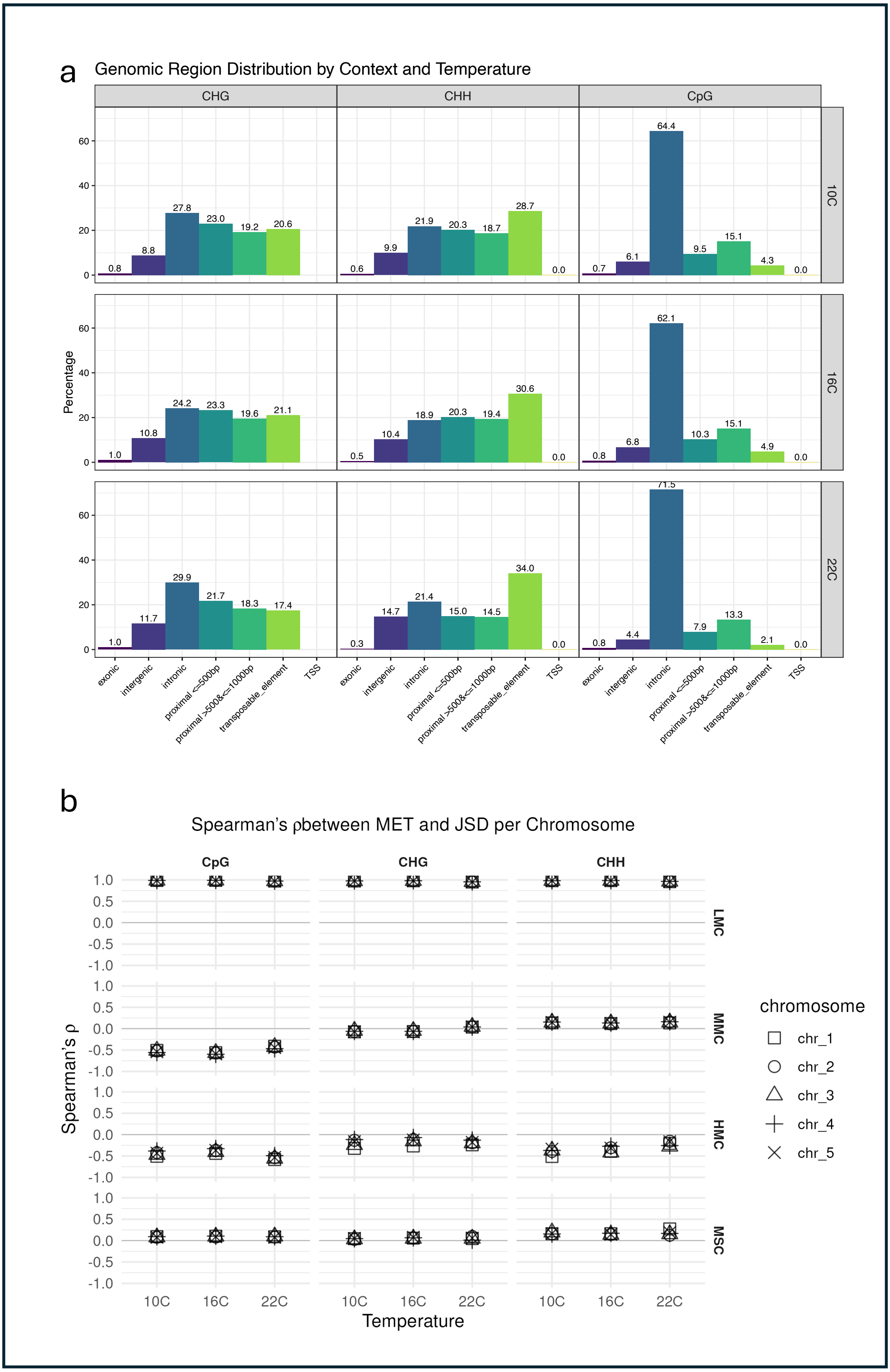

# Figure 4

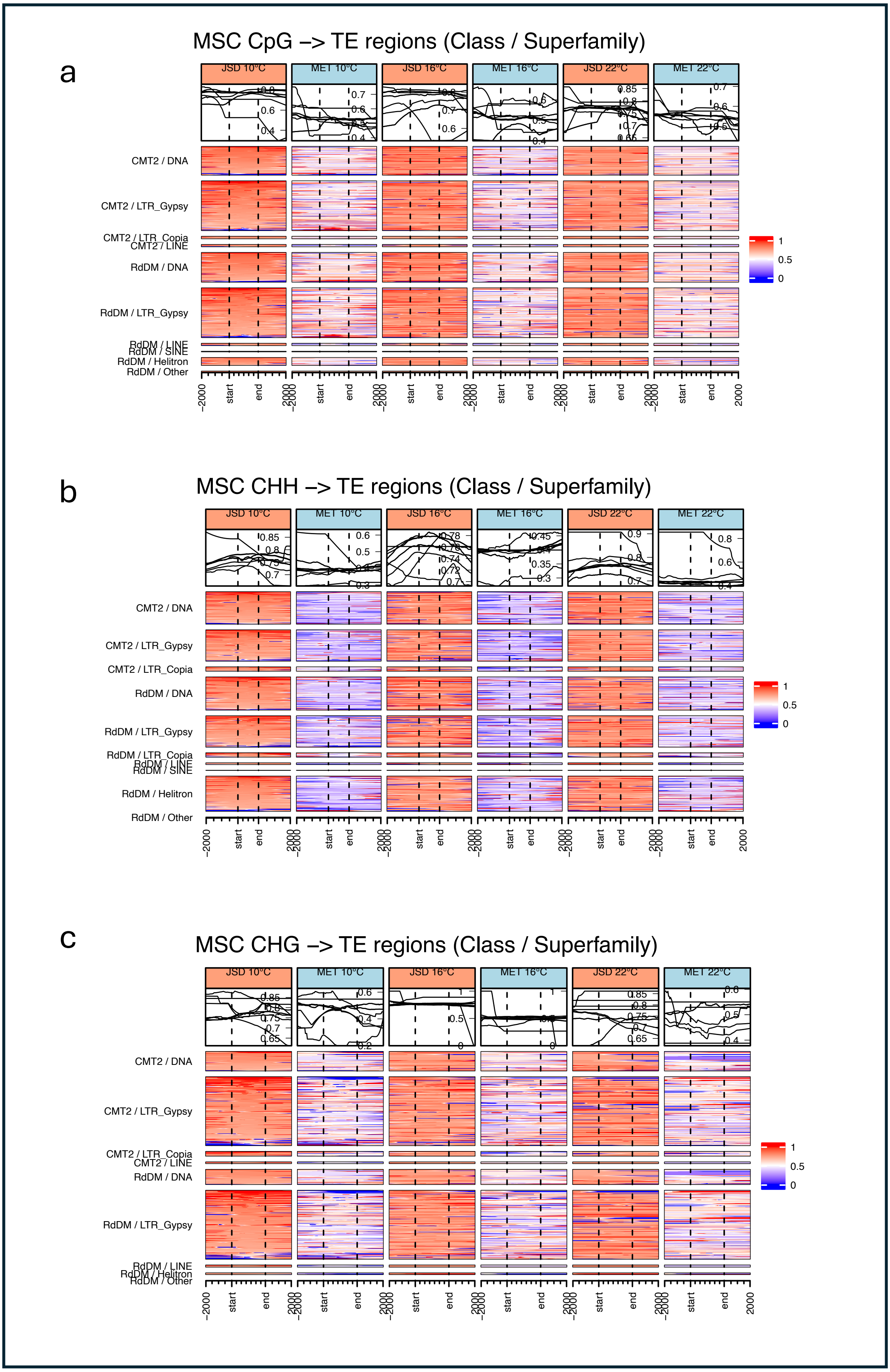
a
MSC CpG –> TE regions (Class / Superfamily)
JSD 10°C
MET 10°C
JSD 16°C
MET 16°C
JSD 22°C
MET 22°C
CMT2 / DNA
CMT2 / LTR_Gypsy
CMT2 / LTR_Copia
CMT2 / LINE
RdDM / DNA
RdDM / LTR_Gypsy
RdDM / LINE
RdDM / SINE
RdDM / Helitron
RdDM / Other
−2000
start
end
2000
1
0.5
0
b
MSC CHH –> TE regions (Class / Superfamily)
CMT2 / DNA
CMT2 / LTR_Gypsy
CMT2 / LTR_Copia
RdDM / DNA
RdDM / LTR_Gypsy
RdDM / LTR_Copia
RdDM / LINE
RdDM / SINE
RdDM / Helitron
RdDM / Other
c
MSC CHG –> TE regions (Class / Superfamily)
CMT2 / DNA
CMT2 / LTR_Gypsy
CMT2 / LTR_Copia
CMT2 / LINE
RdDM / DNA
RdDM / LTR_Gypsy
RdDM / LINE
RdDM / Helitron
RdDM / Other

# Figure 5

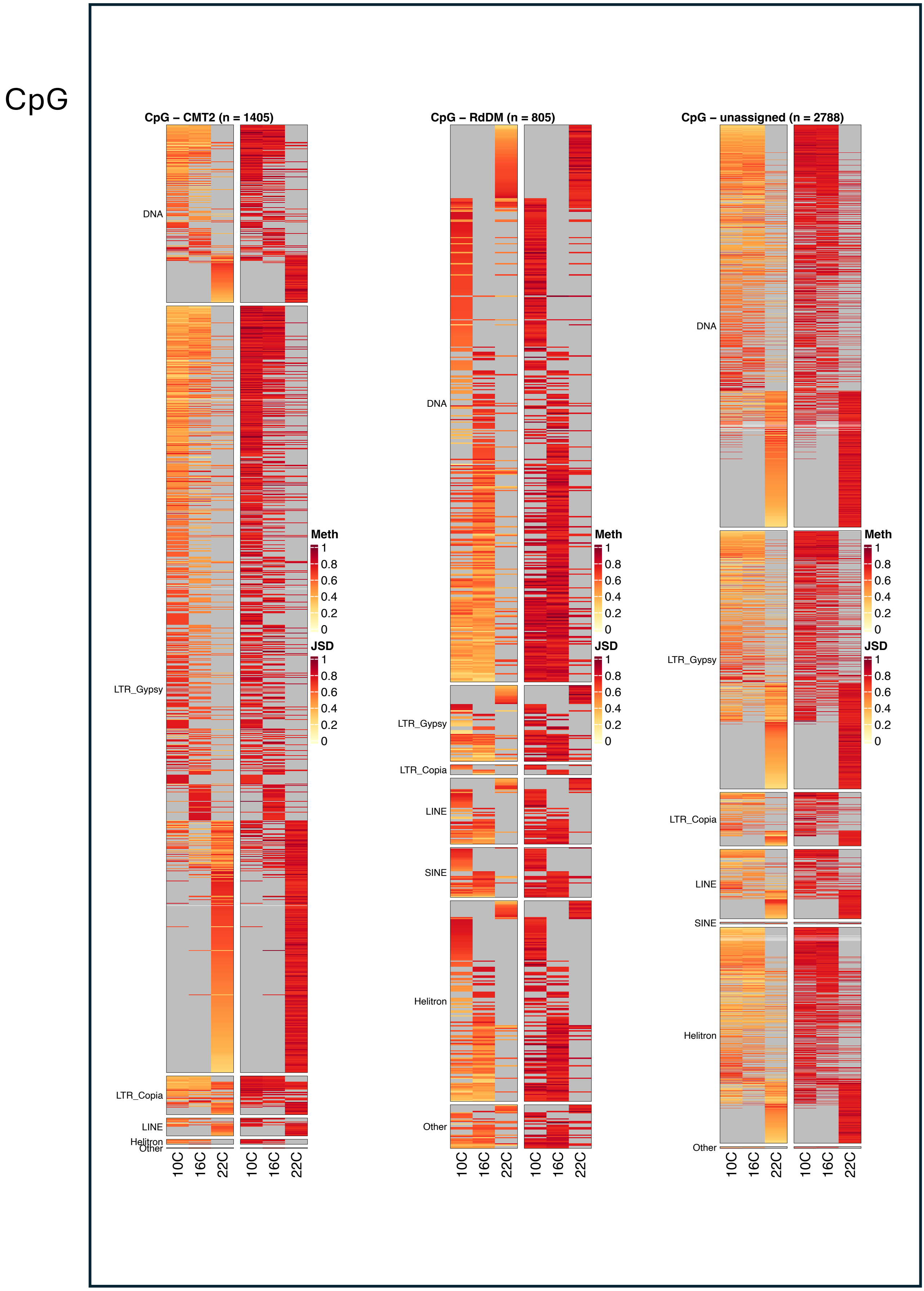

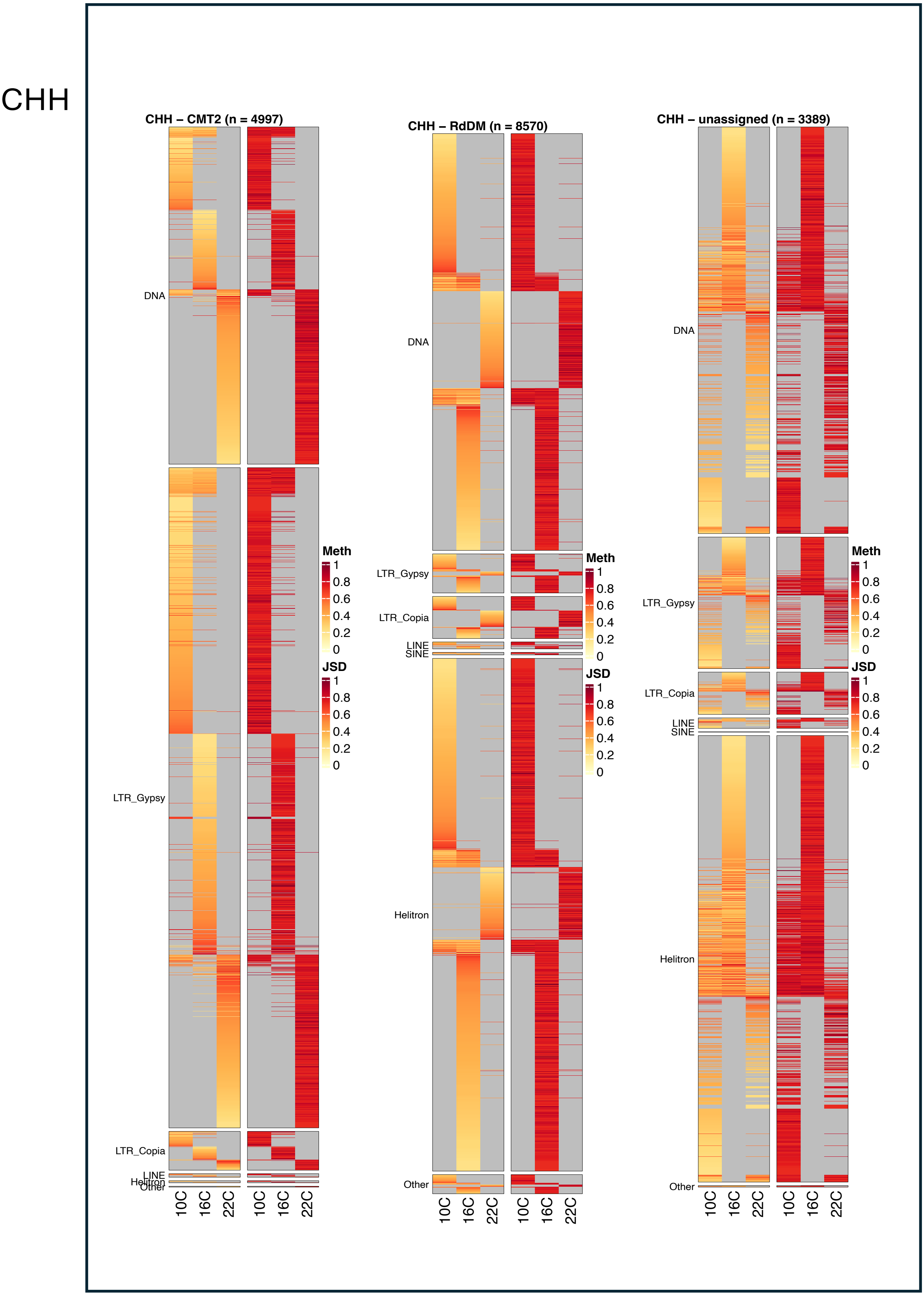

# CHG

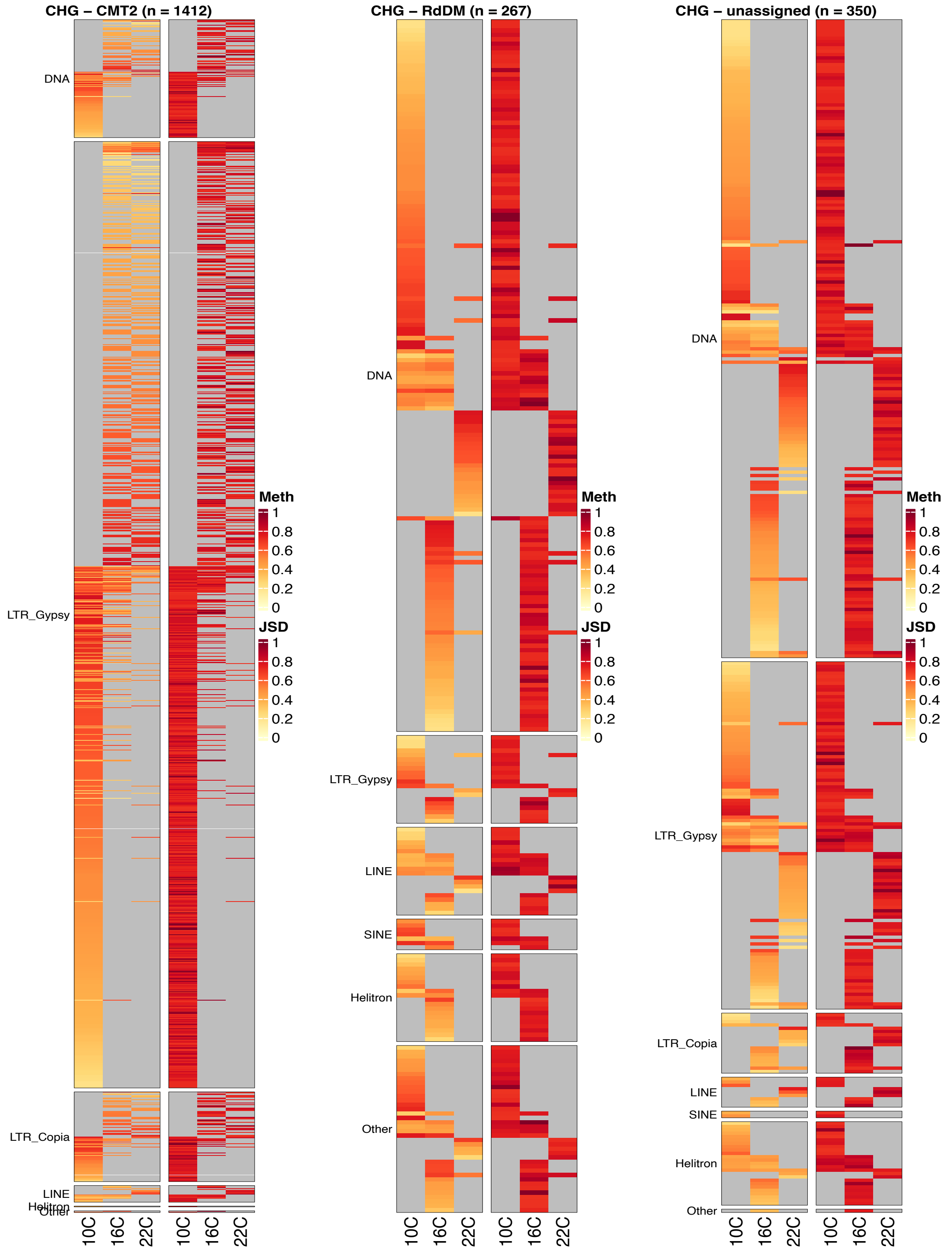

Figure 6

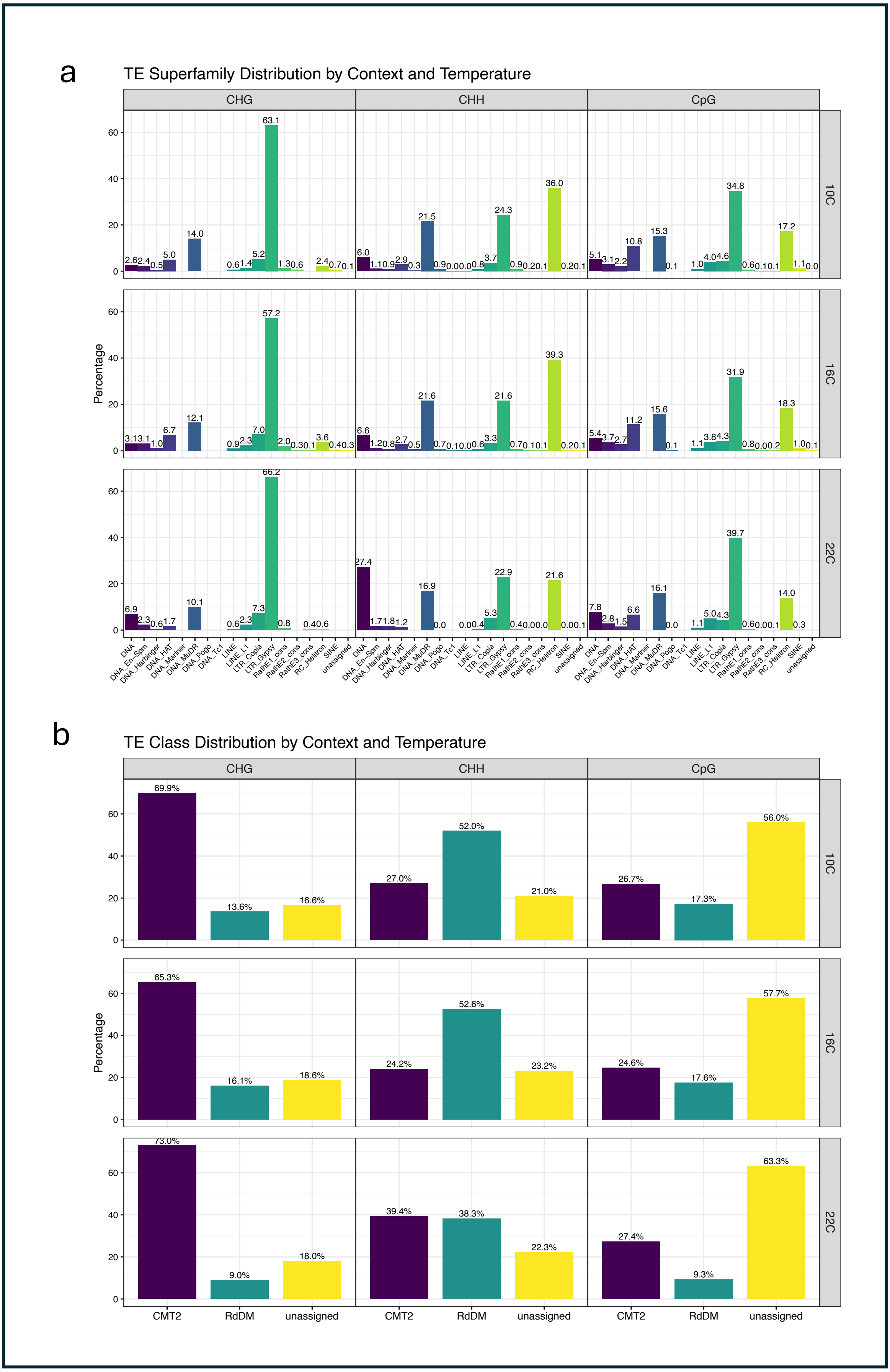

Figure 7

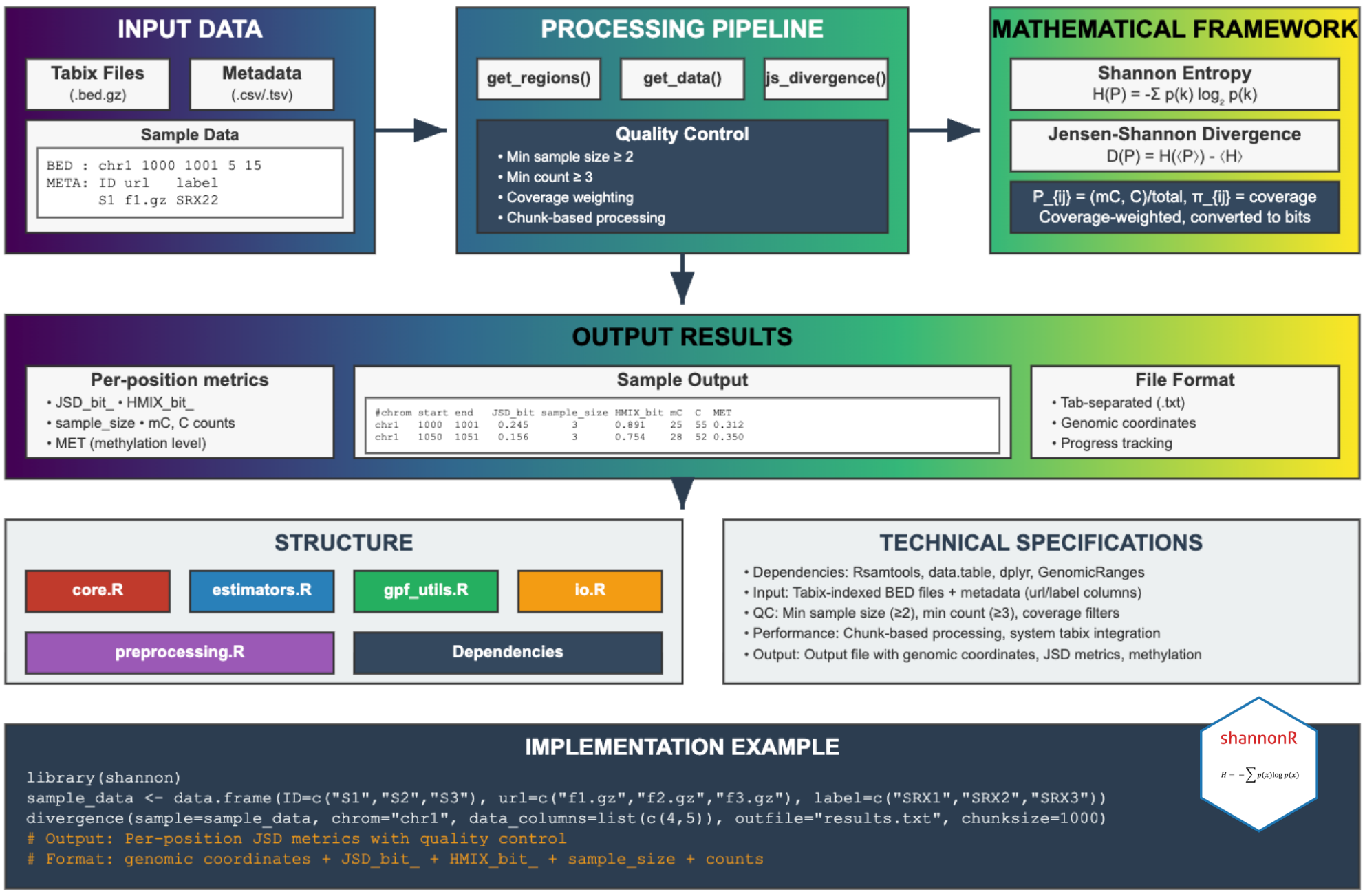

MethDivergence